\documentstyle[psfig]{l-aa}

\newcommand{\Msun}{\mbox{${\rm M}_\odot$}}

\begin{document}

\thesaurus{06(08.05.3; 08.14.1; 08.16.6; 08.02.3)}

\title{Formation of binary millisecond pulsars with relatively high surface dipole magnetic
fields}

\author{
W. Sutantyo \inst{1,2}
\and X.-D. Li \inst{3}
}

\institute{
Department of Astronomy, Institut Teknologi Bandung, Ganesha 10, Bandung 40132,
Indonesia \\(email:sutantyo@bdg.centrin.net.id)
\and  
Astronomical Institute, University of Amsterdam, Kruislaan 403, 1098 SJ
Amsterdam, The Netherlands 
\and           
Department of Astronomy and Astronomical and Astrophysical Center of East China,
Nanjing University, Nanjing 210093, P.R. China (email:lixd@nju.edu.cn)
}

\offprints{W. Sutantyo}

\date{Received April 29; accepted May 11, 2000}

\maketitle

\begin{abstract}
We have carried out numerical evolutionary calculations of binary systems to
investigate the formation of binary millisecond pulsars (pulsars with white
dwarf companions). We apply the ``standard scenario'' in which the binary pulsars
are formed from low-mass and intermediate-mass X-ray binaries as well the
alternative scenario in which the neutron stars are formed by accretion-induced
collapse (AIC) of white dwarfs. The mass transfer processes are carefully followed
by taking into account a number of binary interactions. Assuming that the
magnetic fields of the neutron stars decay due to the accretion, we calculate
the pulsar surface dipole magnetic field strength at the end of the mass
transfer as a function of the final orbital period. We find that while the
observed data of the majority of pulsars are compatible with the derived
relations, we fail to produce binary pulsars with relatively high magnetic fields and short
orbital periods (such as PSR B0655+64). We conclude that those 
systems are most likely formed through common-envelope phase.

\keywords{stars: 	evolution -- 
	  stars:      neutron   -- 
        pulsars:    general  -- 
	  binaries:   general
}

\end{abstract}

\section{Introduction}
It has been suggested that binary pulsars with white dwarf companions (binary
millisecond pulsars or BMSPs) are the descendants of low-mass X-ray binaries
(LMXBs; Joss \& Rappaport 1983, Savonije 1983, and Paczynski 1983) or
intermediate-mass X-ray binaries (IMXBs; Podsiadlowski \& Rappaport 2000; Tauris,  et al. 2000). The white dwarf companions of the pulsars are the remnants of
the donor stars which have dumped matter to the neutron stars during the X-ray
phase. The nearly circular orbits observed in the systems support the view that
extensive mass transfer has happened in the systems (before mass transfer the
orbits must have been eccentric due to the supernova explosions which have formed the
neutron stars, tidal effects become effective to circularize the orbits when the
donor stars swell up to fill up their Roche lobes). Hence, pulsars in those
systems will have accreted some amount of mass and have been recycled (spun up) by the accretion. The
millisecond spin periods observed in many of the pulsars (despite their old
ages, as the nondegenerate companions need some $10^8$ -- $10^9$ yr to become
white dwarfs) indicate that they are indeed recycled pulsars.

Magnetic fields of isolated pulsars do not decay significantly during their
lifetime (Bhattacharya et al. 1992). However, many pulsars which are, or
have been members of binary systems have relatively  weak surface
magnetic fields ($10^8$ -- $10^9$ G).  This leads to the suggestion that the
magnetic field of neutron stars decays due to the accretion (Taam \& van den
Heuvel 1986, Shibazaki et al. 1989) or related effects, such as spin up and spin
down (Srinivasan et al. 1990). However, a number of BMSPs are observed to have
relatively strong fields ($\sim 10^{10}$ -- $10^{11}$ G; see Table 1) which
implies that those neutron stars have accreted only a small amount of matter. We
investigate a number of evolutionary models of binary systems which lead to the
formation of BMSPs and examine whether such high-magnetic field BMSPs can be
formed. In Sect. 2 we discuss the ``standard scenario''  in which the systems
are formed from LMXBs or IMXBs. In Sect. 3 we discuss an alternative scenario in
which the pulsars are formed by accretion-induced collapse (AIC) of white
dwarfs. The discussion and conclusion are given in Sect. 4.

\begin{table}
\caption[ ]{Binary millisecond pulsars with relatively strong magnetic fields.
f(M) is the mass function and $M_{\rm wd}$ is the mass of the white dwarf
companion with the assumption $i = 60 \degr$ (Tauris \& Savonije 1999, Tauris et al. 2000)}
\begin{flushleft}
\begin{tabular}{llllll}
\hline\noalign{\smallskip}
PSR & $P_{\rm orb}$ & $P_{\rm spin}$ & $\log(B)$ &f(M)   & $M_{\rm wd}$ \\
    & (days)        & (ms)           & (G)   &(\Msun)& (\Msun)  \\
\hline\noalign{\smallskip}
B1831-00   & 1.811 & 520.95 & 10.94 &0.000124 &0.075   \\
B0655+64   & 1.029 & 195.67 & 10.07 &0.0714   &0.814 \\
J1232-6501 & 1.863 &  88.3  & 9.97  &0.0014   &0.175 \\
J1157-5112 & 3.507 &  43.6  & $< 9.80$  &0.2546   &$>1.2$  \\
\hline\noalign{\smallskip}
J1803-2712 & 406.781 & 334.42 & 10.89&0.0013  &0.170\\
B0820+02   & 1232.47 & 864.87 & 11.48&0.003   &0.231\\
\noalign{\smallskip}
\hline
\end{tabular}
\end{flushleft}
\end{table}

\section{Binary pulsars formed from LMXBs or IMXBs (``standard scenario'')}
We will first discuss the ``standard'' evolutionary scenario in which BMSPs are
formed from LMXBs and IMXBs. We used an updated version of the numerical stellar
evolution code of Eggleton (1971, 1972, 1973) to keep track the evolution of the
donor star in which the mass-transfer process is carefully followed by taking
into account a number of binary interactions (see Tauris \& Savonije 1999 for
the description of the code). The initial parameters are chosen such as to avoid
spiral-in and common-envelope (CE) evolution (Tauris et al. 2000). We assume
models of systems in which the initial mass of the donor star is between 1 and
4.5 $\Msun$ and the initial orbital period is between 1 -- 20 days. The initial
mass of the neutron star is assumed to be 1.3 $\Msun$. Accretion onto the neutron
star is limited by the Eddington limit ($10^{-8} \ \Msun {\rm{yr}}^{-1}$). Above
this limit, matter which cannot be accepted by the neutron star is ejected
isotropically carrying the specific angular momentum of the neutron star
(``isotropic reemission''). We also take into account the loss of angular
momentum due to gravitational radiation and magnetic braking.  We neglect the
possible mass loss from the accretion disk due to the centrifugal propeller
effect as this effect is not precisely known. We follow the evolution of the
systems starting from the zero-age main-sequence phase of the donor until the
end of the mass transfer phase. We also take into account mass loss prior to the
Roche lobe overflow. The calculations are technically similar to those carried
out by Tauris \& Savonije (1999) and Tauris et al. (2000).

Using population synthesis calculation, Bhattacharya et al. (1992) indicate that
isolated pulsars with high magnetic field do not undergo significant magnetic
field decay during their lifetime ($\sim 10^7$ \ yr). On the other hand, binary
and millisecond pulsars which are members, or have been members of
binaries, have low magnetic field strength ($10^8$ -- $10^9$ \ G). These facts
lead to the conclusion that the magnetic fields of those pulsars decay due to
the interaction in the binaries. More specifically, Taam \& van den Heuvel
(1986) and Shibazaki et al. (1989) suggest that the magnetic field decay is
induced by the mass transfer which has happened in the systems. Cheng \& Zhang
(1998), by assuming the frozen field and incompressible fluid approximations,
derive a relation for the surface dipole magnetic field strength and the amount
of matter accreted by the neutron star:

\begin{equation}
B = \frac{B_f}
{
\left(
1 - C \exp
\left(
- {\frac {\Delta M}{M_{\rm {cr}}}}
\right)
\right)
^{\frac {7}{4}}
}
\end{equation}

\noindent where,

\noindent
$
B_f = 4.3 \times 10^8 \alpha (\dot{M}/\dot{M}_{\rm {Ed}})^{1/2}
{(M/\Msun)}^{1/4} R_6^{-5/4} \\
$

\noindent and,

\noindent 
$
C = 1 - 
\left(
\frac{B_f} {B_{\rm{o}} }
\right)
^{4/7} 
$

\noindent $\alpha$ is an adjustment factor of the order of unity, $\dot{M}_{\rm
{Ed}}$ is the Eddington mass accretion rate, $\Delta M$ is the amount of matter
accreted by the neutron star, $M_{\rm{cr}}$ is the mass of the crust, $R_6$ is the
radius in units of $10^6$ cm, and $B_{\rm{o}}$ is the initial magnetic field
strength of the neutron star. From this formula one can show that a neutron star
with $B_{\rm{o}} \sim 10^{12} {\rm G}$ can decrease its surface magnetic field
to $B \sim 10^{9} {\rm G}$ by accreting only a few hundredth of $\Msun$. 

Using Eq. (1) and $\Delta M$ which we obtained from the evolutionary
calculations, we can derive the surface dipole magnetic field strength of the
neutron stars as a function of the orbital period. The results are shown as the
solid curves in Fig. 1. Here we assume that $\alpha = 0.4$ (this value is chosen
to get the best fit to the observed data based on the eye estimate), $R_6 = 1$,
$\dot{M}/\dot{M}_{\rm {Ed}} = 1$, $M_{\rm{cr}} = 0.1 \ \Msun$,  and $B_{\rm{o}}
=  10^{12}$ G. The results are not sensitive to the values of $B_{\rm o}$ and
$M_{\rm cr}$, but depend on the choice of $\alpha$. The apparent discontinuities
for large donor masses reflect the ``jump'' of the evolutionary timescale when
the donors fill the Roche lobe while crossing the Hertzsprung gap. Also in Fig.
1, we plot the observed data of galactic binary pulsars with circular orbits ($e
< 0.01$) for comparison. We do not include binary pulsars in globular clusters
as they may have had a different history. Note that the majority of pulsars, except a
group of pulsars at the upper left, can be fitted nicely with the curves. This
suggests that the ``standard scenario'' is the most plausible way to form most
systems.

\begin{figure}
%\picplace{5.5 cm}
\psfig{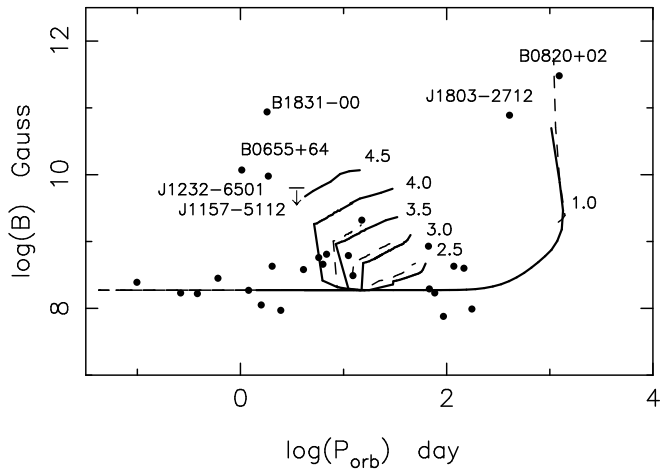}
%\resizebox{\hsize}{!}{\includegraphics{fig_ws.eps}}
\caption{Pulsar magnetic field strength as a function of the orbital period. The
solid curves show the relation for BMSPs formed from LMXBs and IMXBs (``standard
scenario''), the dashed curves indicate the relation for binary pulsars formed
by AIC of white dwarfs (see text for further explanations). The initial mass of
the donor star is shown on each curve. The dots represent the observed binary
pulsars in the galactic disk with circular orbits; for PSR J1157-5112 only the upper limit of $\log(B)$ is shown (we do not include binary pulsars in globular clusters since they may have different evolutionary history).}
\end{figure}

\section{Binary pulsars formed by accretion-induced collapse of white dwarfs}
We will investigate an alternative scenario to form BMSPs in which the neutron
stars are formed by accretion-induced collapse (AIC) of white dwarfs (van den
Heuvel \& Bitzaraki 1995). AIC is expected to happen in systems consisting of a
normal star and a $\sim$ 1.0 -- 1.2 $\Msun$ ONeMg white dwarf (van den Heuvel
1994). If the white dwarf can accrete enough matter to grow its mass up the
Chandrasekhar limit, it will collapse to a neutron star. In this process, about
0.2 $\Msun$ is converted to the binding energy of the neutron star (some amount
of mass may also be ejected), hence the total mass of the system is reduced. As
a result, the orbit somewhat widens and the donor is temporarily detached from
its Roche lobe. Mass transfer is resumed when the donor, which is continuing its
evolution, again fills up its Roche lobe. At this stage the system can be
observed as an LMXB. As in the ``standard scenario'', the system will eventually
end up as a binary pulsar with a white dwarf companion. The difference with the
``standard scenario'' is that, here the donor star is already an evolved star at
the birth of the neutron star (since it must have filled its Roche lobe to
initiate the mass transfer which induces the collapse of the white dwarf), while
in the ``standard scenario'' the donor star is essentially an unevolved star
when the neutron star is born. Since in this scenario the neutron stars are born
when the donors have already been losing mass, one may expect that those neutron
stars will accrete less matter than those formed from the ``standard scenario''.
Consequently, they may have stronger magnetic fields. We will examine this
notion.

Not all accreting white dwarfs can reach the Chandrasekhar limit. Li \& van den
Heuvel (1997) show that there are two groups of systems in which the white dwarf
can grow up the Chandrasekhar limit. One is close binaries with $\sim$ 2 to 3.5
$\Msun$ main-sequence or subgiant companions, and the initial orbital period of
several tenth of a day to several days. The other is wide binaries with low-mass
($\sim$ 1 $\Msun$) red giant companions with long (ten to hundred day) orbital
period. Although the result of Li \& van den Heuvel is aimed for seeking the
progenitors of Type Ia supernovae, it applies to AIC as well. In this work we
take into account these constraints. We assume that the initial mass of the
white dwarf is 1.2 $\Msun$.

During the collapse about 0.2 $\Msun$ ($\Delta M_{\rm {be}}$) is converted to
the binding energy of the neutron stars. Some amount of mass ($\Delta M_{\rm
{ej}}$) may also be ejected from the system.  This will result to the widening
of the system:

\begin{equation}
a = a_{\rm{o}} \frac{M_{\rm {donor}} + M_{\rm {wd}}}{M_{\rm {donor}} + M_{\rm
{ns}}}
\end{equation}

\noindent where $a_{\rm{o}}$ and $a$ are, respectively, the orbital separations
before collapse and after tidal circularization; $M_{\rm {donor}}$ and 
$M_{\rm {wd}}$ are, respectively, the masses of the donor and the white dwarf; 
$M_{\rm {ns}}$ is the mass of the neutron star ($M_{\rm {ns}} = M_{\rm {wd}} - \Delta M_{\rm {aic}}$ where $\Delta M_{\rm {aic}} = \Delta M_{\rm {be}} + \Delta M_{\rm{ej}}$). We assume $\Delta M_{\rm {aic}} = 0.2 \ \Msun$ for all models. We neglect any kick velocity which may be received by the neutron star due to asymmetric process during the collapse.

We keep track of the evolution of the systems by following the accretion to the
white dwarf and subsequently to the neutron star. The results are shown as the
dashed curves in Fig. 1 and we find that those are essentially the same as those
of the ``standard scenario''. This is due to the fact that to maintain
relatively high surface magnetic field ($\ga 10^9$ G) the neutron star must
accrete only a small fraction (at most a few percent) of the transferred mass,
so this condition is not affected so much by the case whether the neutron star
already exists at the beginning of the mass transfer or it exists somewhat later
when the donor is already losing some amount of mass. Van den Heuvel \&
Bitzaraki (1995) suggest that PSR B1831-00 is a candidate of system formed from
this scenario. But, as is evidenced from Fig. 1, our results seem to rule out
this possibility.

\section{Discussion and conclusion}
We have derived the relations between the surface magnetic field strengths and the
final orbital periods of BMSPs. The observed data of BMSPs mostly fit well with
the derived relations. This is compatible with the view that those systems are
indeed the descendants of LMXBs or IMXBs. The existence of high-magnetic field
pulsars with long orbital periods (PSR B0820+02 and J1803-2712) can naturally be
explained that they have originated from long period LMXBs. In such long period
systems the red giant donors have a deep convective envelope at the onset of the
mass transfer, hence the mass transfer happens violently (super Eddington) and
occurs in short timescale (van den Heuvel 1994) such that the neutron stars can
only accrete a small fraction of the transferred mass. As a result, the pulsars
only experience a mild decay of the magnetic field.

We, however, fail to produce high-magnetic field pulsars with short orbital
periods such as PSR B0655+64 and PSR B1831-00. We conclude that those systems
cannot be formed from either ``standard'' or AIC scenario. Tauris et al. (2000)
also reach similar conclusion that binary pulsars with high-mass CO/ONeMg white
dwarf companions (PSR B0655+64 belongs to this category as it has a $\sim 0.8
\Msun$ companion) and short orbital periods cannot be formed from LMXBs or IMXBs.
Then, the most plausible scenario for forming those systems is that they are
formed from the common-envelope (CE) evolution as is originally suggested by van
den Heuvel \& Taam (1984) for PSR B0655+64. In this scenario, the initial mass
of the companion star is considerably larger than the mass of the neutron star
(say $5 \Msun$) and the initial orbital period is very long ($\sim 1000$ days).
The companion will fill up its Roche lobe when it ascends the asymptotic giant
branch (AGB), i.e. after helium exhaustion in its core. Mass transfer from such
an evolved and massive companion tends to be unstable. Such a system undergoes a
common envelope (CE) phase in which the neutron star spirals into the envelope
of the primary and both stars are embedded in a common envelope. During the
spiral-in of the neutron star, the envelope is blown off at the expense of the
orbital energy (and probably also in combination with some other energy sources;
i.e., accretion energy, recombination energy etc.). At the end of the CE phase,
the system consists of a $\sim 1 \Msun$ CO white dwarf (initially the CO core of
the primary) and a neutron star. As a large fraction of the orbital energy is
used to expel the envelope, the orbital separation of the system is very small
($P_{\rm orb} \la 1 {\rm d}$). The system resembles closely PSR B0655+64. Since
the duration of the CE phase is short ($\sim 10^3$ -- $10^5$ yr), the neutron
star will not accrete much matter, so it will maintain its high magnetic field.
However, Chevalier (1993) and Brown (1995) have suggested that a spiral-in
neutron star may experience a hypercritical accretion to become a black hole.
Despite this suggestion, we argue that the existence of short orbital period
BMSPs with relatively strong dipole fields indicates that CE evolution is still
an attractive way to form binary pulsars. 

The remaining problem is that, the small mass function of PSR B1831-00 may imply
a small mass of the white dwarf ($\sim 0.075 \Msun$ if we assume $i = 60
\degr$). This mass is much too small to be reconciled with the CE evolution. In
this respect, the small mass function may indicate, instead, a small inclination
angle. However, if we require $M_{\rm wd} \ga 0.6 \Msun$ then $i \la 7.5 \degr$.
Assuming a random orientation of the orbit, the probability of observing a
binary with such a small inclination angle is only 0.8\%. Another possibility is
that the system may result from AIC of a 1+1.2 $\Msun$ main-sequence and white
dwarf system with unstable mass transfer (Li \& Wang 1998). But the outcome is
highly speculative. We conclude therefore that the history of PSR B1831-00 is
indeed very special and it needs further studies.

We have used a specific relation between magnetic field strength B and
amount of matter accreted by the neutron star $\Delta M$, as given in Eq. (1).
This relation well reproduces the observed relation between
B and orbital period of BMSPs originating from LMXBs
and IMXBs. This is certainly a strong argument in favour of a relation of
the type of Eq. (1), however it is not necessarily a confirmation for this
type of relation. In the literature, there are at least two other relations
have been suggested (see e.g. Shibazaki et al. 1989 and Urpin \& Geppert,
1995). The field decay described by these relations are more slow or rapid
compared to that by Eq. (1). But, evidently, they will give similar ``well
fitting''for the vast majority of the BMSPs (Li \& Wang,
1998) in which PSR B0655+64 and PSR B1831-00 will always remain the
exceptions.

\begin{acknowledgements}
We would like to thank E. van den Heuvel for stimulating discussions and advice on the manuscript, and to G. Savonije, T. Tauris and J. Dewi for
stimulating discussions. We also thank R. Ramachandran for providing us with
the data of binary pulsars. W.S. gratefully acknowledges the financial supports
from the NWO Spinoza Grant 08-0 to E.P.J. van den Heuvel, and the Leids
Kerkhoven-Bosscha-Fonds which have enabled him to stay at the Astronomical
Institute, University of Amsterdam. X.L. was supported by National Natural
Science Fundation of China and National Major Project for Basic Research of
China.
\end{acknowledgements}


\begin{thebibliography}{}

\bibitem{} Bhattacharya D., Wijers R.A.M.J., Hartman J.W., Verbunt F. 1992,
      A\&A 254, 198

\bibitem{} Brown G.E. 1995,
      ApJ 440, 270

\bibitem{} Cheng K.S., Zhang C.M. 1998,
      A\&A 337, 441

\bibitem{} Chevalier R.A. 1993,
      ApJ 411, L33

\bibitem{} Eggleton P.P 1971,
      MNRAS 151, 351

\bibitem{} Eggleton P.P 1972,
      MNRAS 156, 361

\bibitem{} Eggleton P.P 1973,
      MNRAS 163, 279

\bibitem{} Joss P.C., Rappaport S.A. 1983,
      Nat 304, 419

\bibitem{} Li X.-D., van den Heuvel E.P.J. 1997,
      A\&A 322, L9

\bibitem{} Li X.-D., Wang Z.-R. 1998,
      ApJ 500, 935

\bibitem{} Paczynski B. 1983,
      Nat 304, 421

\bibitem{} Podsiadlowski P., Rappaport S.A. 2000,
      ApJ 529, 946

\bibitem{} Savonije G.J. 1983,
      Nat 304, 422

\bibitem{} Shibazaki N., Murakami T., Shaham J., Nomoto K. 1989,
      Nat 342, 656

\bibitem{} Srinivasan G., Bhattacharya D., Muslimov A.G., Tsygan A.I. 1990,
      Current Science 59, 31

\bibitem{} Taam R.E., van den Heuvel E.P.J. 1986,
      ApJ 305, 235

\bibitem{} Tauris T.M., Savonije G.J. 1999,
      A\&A 350, 928
 
\bibitem{} Tauris T.M., van den Heuvel E.P.J., Savonije G.J. 2000,
      ApJ 530, L93

\bibitem{} Urpin V., Geppert U., 1995
      MNRAS 275, 1117

\bibitem{} van den Heuvel E.P.J. 1994,
      in: Interacting binaries,
      eds. \ S.N. Shore, M. Livio, E.P.J. van den Heuvel,
      Springer Verlag, Berlin, p. \ 263

\bibitem{} van den Heuvel E.P.J., Bitzaraki O. 1995,
      A\&A 297, L41

\bibitem{} van den Heuvel E.P.J., Taam R.E. 1984,
      Nat 309, 235

\end{thebibliography}
\end{document}